\documentclass[reprint,10pt,
prl,
aps, 
amsmath,amssymb,
superscriptaddress,
floatfix
]{revtex4-2}

\usepackage{circuitikz}
\usepackage{siunitx}
\usepackage{multirow}
\usepackage{braket}

\begin{document}
\newcommand{\Emptyset}{\text{\o}}

\title{Tomographic Signatures of Interacting Majorana and Andreev States in Superconductor-Semiconductor Transmon Qubits}

\author{Daniel Dahan}
\affiliation{Department of Physics, Ben-Gurion University of the Negev, Beer-Sheva 84105, Israel}

\author{Konstantin Yavilberg}
\affiliation{Department of Physics, Ben-Gurion University of the Negev, Beer-Sheva 84105, Israel}

\author{Talya Shnaider}
\affiliation{Department of Physics, Ben-Gurion University of the Negev, Beer-Sheva 84105, Israel}

\author{Elena Lupo}
\altaffiliation{Current address: Quantum Computing Analytics (PGI-12),
Forschungszentrum J\"{u}lich, 52425 J\"{u}lich, Germany}
\affiliation{Advanced Technology Institute and School of Mathematics and Physics, University of Surrey, Guildford, GU2 7XH, UK}

\author{Malcolm R. Connolly}
\affiliation{Blackett Laboratory, Imperial College London, South Kensington Campus, London SW7 2AZ, United
Kingdom}

\author{Eran Ginossar}
\affiliation{Advanced Technology Institute and School of Mathematics and Physics, University of Surrey, Guildford, GU2 7XH, UK}

\author{Eytan Grosfeld}
\affiliation{Department of Physics, Ben-Gurion University of the Negev, Beer-Sheva 84105, Israel}

\begin{abstract}
    Semiconductor-based Josephson junctions embedded within a Cooper-pair-box can host complex many-body states, such as interacting Andreev states and potentially other quasi-particles of topological origin. Here, we study the insights that could be revealed from a tomographic reconstruction of the Cooper-pair charge distribution of the junction prepared in its ground state. We posit that  interacting and topological states can be identified from distinct signatures within the probability distribution of the charge states. Furthermore, the comprehensive dataset provides direct access to information theory metrics elucidating the entanglement between the charge sector of the superconductor and the microscopic degrees of freedom in the junction. We demonstrate how these metrics serve to further classify differences between the types of excitations in the junction.
\end{abstract}

\date{\today}

\maketitle

\emph{Introduction}.--- Superconductor-semiconductor junctions exhibit intriguing excitations like Andreev states and can potentially host elusive Majorana zero modes. When the superconductors form isolated islands, the interplay between phase-dependent Andreev reflection and charging energy generates collective states comprising correlated superconducting charge states and microscopic low-lying excitations specific to the junction. The resulting states admit properties potentially conducive to processing quantum information~\cite{zazunov2003andreev,yavilberg2015fermion,lupo2022implementation, Pita-Vidal2023, PhysRevLett.124.246803}; moreover, coupling the junction to a microwave cavity enables spectroscopic methods for extracting state properties \cite{ginossar2014microwave, yavilberg2019differentiating, doi:10.1126/science.aab2179, PhysRevResearch.2.033493, PhysRevLett.131.097001, PRXQuantum.3.030311, PhysRevB.104.174517, PhysRevLett.121.047001, Scarlino2019,haver2023electromagnetic}. While spectroscopy is a sensitive tool for probing the spectrum and transition rules of a junction, a more comprehensive characterization of the junction's microscopic states remains a high concern.

The study of correlations across partitions of quantum systems offers a unique lens into topological~\cite{kitaev2006topological,levin2006detecting} and interacting phases. Superconductor-semiconductor junctions naturally divide into two subsystems: the condensate, comprising the bosonic Cooper pairs, and the semiconductor, hosting low-lying fermionic excitations in the junction. The superconducting charge states describe the first subsystem that readily couples to the electromagnetic fields in the cavity. This prompts a compelling query: can insights into the junction be gleaned solely from observing the superconducting charge states? Specifically, can the charge composition of these correlated states unveil information about the existence and nature of fermionic excitations in the junction? 

In this paper, we study the structure of charge states in the superconductor-semiconductor junction  and the potential insights it can provide into the weak link's underlying low-lying fermionic states in various scenarios. We will leverage a charge-basis reduced density matrix as our primary tool, which comprises the full set of data that could be reconstructed using the interactions with a probe circuit~\cite{lupo2024proposal}. Our focus lies in exposing fingerprints of the fermionic low-lying states within this data. These include the diagonal elements of the density matrix, describing the probability per charge state, and the entanglement entropy as a measure for the entanglement of the fermionic degrees of freedom in the junction with the charge degrees of freedom.

Specifically, we investigate a Transmon coupled to a weak link, such as an interacting dot or a short semiconducting wire \cite{casparis2018superconducting, PhysRevLett.115.127002}. The superconducting islands comprising the Transmon give rise to quantized plasma oscillations of Cooper pairs, which interact with the fermionic states upon their injection into and out of the dot. These interactions endow the bosonic plasma oscillations with additional structure coming from the low-lying excitations of the junction; we shall dub such resonances ``interacting Andreev states''. In the case of topological superconductivity, Majorana states further dress the dot states. The resulting interacting Andreev and Majorana states can be challenging to disentangle, making it crucial to pinpoint the pertinent correlations that could distinguish them.

\begin{figure}[t]
    \includegraphics[width=0.9\linewidth]{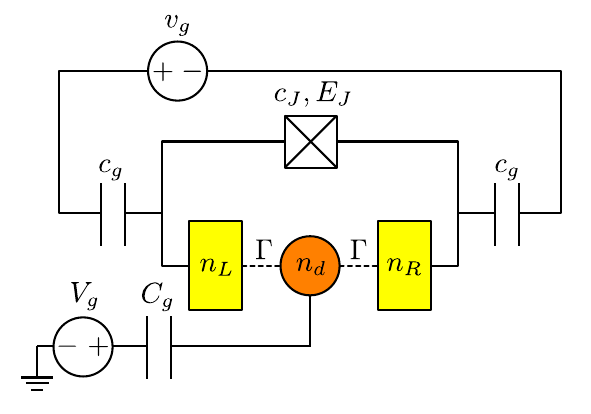}
	\caption{Quantum circuit describing the qubit considered. Two superconductors (yellow) with $n_L$, $n_R$ cooper pairs, are coupled via a dot (red) with $n_d$ fermions, forming a Josephson junction. It is connected in parallel to another Josephson junction, with Josephson energy $E_J$, and is controlled by two gates, inducing offset charges $n_g=c_g v_g/(2\mathrm{e})$ and $N_g=V_g C_g/(2\mathrm{e})$, where $\mathrm{e}$ is the electron charge. 
    \label{fig:circuit}}
\end{figure}

\emph{Charge-basis density matrix of a single, interacting Andreev state}.---
We start with the model for the ``Dot-Transmon'' (DT), a prototypical effective model consisting of the charge and phase dynamics of a modified Transmon and its interactions and charge exchange with a normal dot (the latter can also represent the low energy dynamics of a short wire). For convenience, we divide the full Hamiltonian into two parts, $\hat{H} = \hat{H}_{\text{T}}+\hat{H}_{\text{D}}$. The first, the modified Transmon Hamiltonian, accounts for the Transmon's dynamics as well as for the dot's charging energy and interaction with a nearby gate,
\begin{align}
\label{eq: dot-Transmon H_T}
\hat{H}_\text{T} = 4E_C(\hat{n} - n_g)^2 + E_C'\left(\hat{N}-N_g\right)^2 -E_J \cos \hat{\varphi}.
\end{align}
The second is responsible for the interaction of the dot's electrons with the magnetic field and with the nearby superconductors via the proximity effect,
\begin{align}
\label{eq: dot-Transmon H_dot}
&\hat{H}_{\text{D}} = B(\hat{c}^\dagger_\uparrow \hat{c}_\uparrow - \hat{c}^\dagger_\downarrow \hat{c}_\downarrow) 
 \nonumber \\
&+2\Gamma \cos (\hat{\varphi}/2) \left(e^{i\hat{\delta}}\hat{c}_\uparrow \hat{c}_\downarrow + e^{-i\hat{\delta}}\hat{c}^\dagger_\downarrow \hat{c}^\dagger_\uparrow \right).
\end{align}
Here $\hat n=\frac12(\hat{n}_L-\hat{n}_R)$ is the relative number of Cooper pairs between the islands (with $\hat{n}_{L/R}$ the number on the left/right island) and is conjugate to the phase difference $\hat{\varphi} =\hat{\varphi}_L-\hat{\varphi}_R$ (with $\hat{\varphi}_{L/R}$ the phase on the left/right island); $\hat{N} = \hat{n}_L+\hat{n}_R$ counts the total number of Cooper pairs exceeding neutrality in the superconducting islands and is conjugate to the average phase $\hat{\delta}=\frac12(\hat{\varphi}_L+\hat{\varphi}_R)$; $n_g$ ($N_g$) is controlled by a side (top) gate; $E_C$ and $E_J$ are the charging and Josephson energy scales associated with the Transmon (we assume that $E_J$ is predominantly set by a parallel Josephson junction, neglecting the dot's contribution, see Fig.~\ref{fig:circuit}); $E_C'$ controls the energy of a given charge occupation of the dot, with the phase $\hat{\delta}$ serving as the generator for charge transfer into the dot. The operator $c^\dagger_\alpha$ creates a fermion on the dot with spin $\alpha \in \{\uparrow, \downarrow\}$; $B$ represents its Zeeman splitting and $\Gamma$ its induced pairing. 
 
To solve the model it is advantageous to span our space by the four possible dot occupation states $\sigma \in \left\{ |\Emptyset\rangle, \left|\uparrow\right\rangle,\left|\downarrow\right\rangle, \left|\uparrow\downarrow\right\rangle \right\}$. Since our system conserves the total number of particles $N_t$ (with each Cooper pair contributing $1$ and each fermion contributing $1/2$), the dot's occupation $n_d$ is locked to the number of Cooper pairs in the islands $N$ via $N_{t}=N+n_d/2$. We can further simplify the model by focusing on an even $n_d$, being one of the allowed sectors, with the resulting Hamiltonian 
\begin{align}
\label{eq: dot-Transmon H_dot matrix}
    H=\left(\begin{array}{cc}
	H_{\text{T}}^{(N)}[n_g, N_g] &-2\Gamma \cos (\hat{\varphi}/2)  \\
	-2\Gamma \cos (\hat{\varphi}/2) & H_{\text{T}}^{(N-1)}[n_g,N_g]
\end{array} \right),
	\end{align}
operating on $\mathbf{f}(\varphi)=\left(f_{\Emptyset}(\varphi), f_{\uparrow\downarrow}(\varphi)\right)^{\intercal}$, with $f_\sigma(\varphi)=\langle\varphi|\sigma\rangle$ the wavefunction associated with the occupation $\sigma$. The notation $H^{(N)}_\text{T}[n_g, N_g]$ represents a projection of $H_{\text{T}}$ on a specific eigenvalue of $\hat{N}$. While $n_L, n_R\in\mathbb{Z}$ in this case, the quantum numbers $N$ and $n$ are constrained by the fermionic parity \cite{PhysRevB.103.245308, Lee2014, PhysRevLett.129.227701} in the system and by the value of $N_t$, as summarized in Table~\ref{table:dot-Transmon}. Notice that the last two conditions, \emph{i.e.}, the quantization of $n$ and periodicity (or anti-periodicity) of the wavefunctions in $\varphi$, are equivalent to each other, as reflected in the charge basis decomposition of the wavefunctions $f_\sigma(\varphi) = \sum_{n} e^{i\varphi n} A_{\sigma,n}$. From now on, in order to simplify the representation of the results we will consider only the case $N_t = 0$, \emph{i.e.}, $\hat{N}$ is constrained to the two values $N=0, -1$.  
%%%%%%%%%%%%%%%%%%%%%%%%%%%%%%%%%%%%%%%%%%%%%
\begin{table}[b]
\begin{tabular}{|c||c|c|}
\hline
 & $f_{\Emptyset}(\varphi)$ & $f_{\uparrow\downarrow}(\varphi)$\\
 \hline
$n_d=$ & $0$ & $2$  \\[0.1cm]
$N=$ & $N_{t}$ & $N_{t}-1$   \\[0.1cm]
$n\in$ & $\mathbb{Z}+\frac{1-(-1)^{N_t}}{4}$ & $\mathbb{Z}+\frac{1+(-1)^{N_t}}{4}$ \\[0.1cm]
$\varphi\rightarrow \varphi+2\pi$
& $(-1)^{N_t}f_{\Emptyset}(\varphi)$ & $(-1)^{N_t+1}f_{\uparrow\downarrow}(\varphi)$ \\[0.1cm]
\hline
\end{tabular}
\caption{Charge state composition and periodicity of the two states, $f_{\Emptyset}(\varphi)$ and $f_{\uparrow\downarrow}(\varphi)$, associated with an empty and filled dot: the dot occupation (first line), the total number of Cooper pairs in the superconducting islands (second line), the quantization of the relative charge $n$ (third line), and the periodicity of the wavefunctions under $\varphi\to\varphi+2\pi$ (fourth line). \label{table:dot-Transmon}}
\end{table}
%%%%%%%%%%%%%%%%%%%%%%%%%%%%%%%%%%%%%%%%%%%%%

\begin{table*}[t] 
    \begin{tabular}{|c||c|c|c|c|c|c|c|c|}
    \hline
    & $f_{e,e,\Emptyset}(\varphi)$ 
    & $f_{e,e,\uparrow\downarrow}(\varphi)$ 
    & $f_{o,o,\Emptyset}(\varphi)$ 
    & $f_{o,o,\uparrow\downarrow}(\varphi)$ 
    & $f_{o,e,\uparrow}(\varphi)$ 
    & $f_{o,e,\downarrow}(\varphi)$ 
    & $f_{e,o,\uparrow}(\varphi)$ 
    & $f_{e,o,\downarrow}(\varphi)$\\
    \hline
    $n_d = $ & $0$ & $2$ & $0$ & $2$ &  $1$ & $1$ &  $1$ & $1$  \\[0.1cm]
    $N=$ & $N_t$ & $N_t-1$ & $N_t$ & $N_t-1$ &  $N_t-\frac{1}{2}$ & $N_t-\frac{1}{2}$ &  $N_t-\frac{1}{2}$ & $N_t-\frac{1}{2}$  \\[0.1cm]
    $n\in$ & $\mathbb{Z}$ & $\mathbb{Z}+\frac{1}{2}$ & $\mathbb{Z}+\frac{1}{2}$ & $\mathbb{Z}$ 
    & $\mathbb{Z}+\frac{1}{4}$  & $\mathbb{Z}+\frac{1}{4}$ & $\mathbb{Z}-\frac{1}{4}$ & $\mathbb{Z}-\frac{1}{4}$\\[0.1cm]
    $\varphi\rightarrow \varphi+2\pi$
    & $f_{e,e,\Emptyset}(\varphi)$ 
    & $-f_{e,e,\uparrow\downarrow}(\varphi)$ 
    & $f_{o,o,\Emptyset}(\varphi)$ 
    & $-f_{o,o,\uparrow\downarrow}(\varphi)$ 
    & $if_{o,e,\uparrow}(\varphi)$ 
    & $if_{o,e,\downarrow}(\varphi)$ & 
    $-if_{e,o,\uparrow}(\varphi)$ 
    & $-if_{e,o,\downarrow}(\varphi)$
    \\
    \hline
    \end{tabular}
    \caption{Charge state composition and periodicity of states of the interacting Majorana-Andreev system: the dot occupation (first line), the total number of Cooper pairs in the superconducting islands (second line), the quantization of the relative charge $n$ (third line), and the periodicity of the wavefunctions under $\varphi\to\varphi+2\pi$ (fourth line). \label{eq: boundary conditions table - majorana dot Transmon}}
\end{table*}

Due to the induced pairing interaction in the dot, each Transmon level is split into a doublet separated by $\sim\Gamma$. Assuming that $\Gamma\ll \omega_p$ with $\omega_p=\sqrt{8E_C E_J}$ the plasma frequency, we can project the Hamiltonian to the Transmon's ground state, whose energy is $\epsilon_{\text{T},0}(n_g)=\epsilon_0 + (-1)^{n_d/2}t_0\cos(2\pi n_g)$ with dispersion $t_0=-16E_C\sqrt{2/\pi} \left(E_J/2E_C\right)^{3/4}\exp\left(-\sqrt{8E_J/E_C}\right)$ \cite{koch2007charge}. In order to find an explicit form for the pairing term in Eq.~(\ref{eq: dot-Transmon H_dot matrix}) we take the ground state wave function in the harmonic approximation, namely $f_{\Emptyset}(\varphi),f_{\uparrow\downarrow }(\varphi) \simeq \psi(\varphi)$. Here $\psi(\varphi)=\left(\pi^{1/4}\sqrt{\varphi_0}\right)^{-1}\exp\left(-\varphi^2/2\varphi_0^2\right)$ with $\varphi_0=(8E_C/E_J)^{1/4}$. In this approximation, the wavefunction's periodicity is neglected and the limits of integration can be extended to $\pm \infty$. This leads to
\begin{align}\label{eq: dot-Transmon hamiltonian gauged and projected}
    H\simeq \varepsilon_0(N_g)+\varepsilon_x \sigma_x + \varepsilon_z(n_g,N_g) \sigma_z.
\end{align}
Here $\varepsilon_x= -2\Gamma \exp(-\varphi_0^2/16)$, $\varepsilon_z(n_g,N_g)= -E_C'(N_g+ \frac{1}{2})+ t_0\cos(2\pi n_g)$, $\varepsilon_0(N_g)=\epsilon_0+E_C'(N_g^2 +N_g +\frac{1}{2}) $ 
%{\color{blue}{[Corrected sign]}} 
and $\sigma_x$ and $\sigma_z$ are the Pauli matrices operating in the dot's occupation space $|\Emptyset\rangle$, $\left|\uparrow\downarrow\right\rangle$ \footnote{Note that in the case of an asymmetric dot-island coupling, i.e. a different $\Gamma$ on each side of the junction, the Transmon's ground and first excited state remain coupled. This can be solved perturbatively but we will not address this here.}. The eigenvalues of Eq.~(\ref{eq: dot-Transmon hamiltonian gauged and projected}) are
\begin{equation}
\varepsilon_\pm = \epsilon_0 +
E_C'\left(N_g^2+N_g+\frac{1}{2}\right)  \pm 
\sqrt{ \varepsilon_x^2+\varepsilon_z^2(n_g,N_g) },
\end{equation}
and the eigenstates 
$|+\rangle=\cos(\theta)|\Emptyset\rangle+\sin(\theta)\left|\uparrow\downarrow\right\rangle$, and $|-\rangle=\sin(\theta)|\Emptyset\rangle-\cos(\theta)\left|\uparrow\downarrow\right\rangle$, where $\tan(2\theta)= \varepsilon_x/\varepsilon_z(n_g,N_g)$. %{\color{blue}{[corrected]}}. 
The density matrix for the ground state is $\rho_\text{gs} = |-\rangle\langle -|$, or explicitly:
\begin{align}
\rho_\text{gs}=	\left(\begin{array}{cc}
\sin^2(\theta) &-\frac{1}{2}\sin(2\theta) \\ -\frac{1}{2}\sin(2\theta)&\cos^2(\theta) 
\end{array} \right).
\end{align}
From here we can extract the entanglement entropy via $S = -\text{Tr}\rho_{[\tau]} \log \rho_{[\tau]}$, where $\rho_{[\tau]}$ is the reduced density matrix of the ground state obtained after tracing out the dot and the total-charge degrees of freedom. The full calculation is presented in the Supplementary Material \ref{sec: appendix dot-Transmon} and gives the result:
\begin{align}
S=-\sin^2\left(\theta \right)\log\sin^2\left(\theta\right)-\cos^2\left(\theta\right)\log\cos^2\left(\theta \right).
\end{align}
The maximally entangled state, \emph{i.e.}, the equal amplitude superposition of $|\Emptyset\rangle$ and $\left|\uparrow\downarrow\right\rangle$, occurs for $N_g=N_{g,\text{max}}$,
\begin{align}
N_{g,\text{max}} = \frac{t_0}{E_C'} \cos(2\pi n_g)- \frac{1}{2},
\end{align}
where the entanglement entropy attains its maximal value, $S_\text{max} = \log(2)$, while concurrently, the dot gets occupied (on average) by a single fermion.

\emph{Charge-basis density matrix of coupled, interacting Andreev and Majorana states}.--- Next, we introduce Majorana fermions to the DT analyzed in the previous section. This is achieved by including in the Hamiltonian the term \cite{yavilberg2019differentiating}:
\begin{align}
\label{eq: majorana-dot-Transmon H_gamma}
H_{\text{M}} = w e^{i\hat{\delta}/2 } \left[ie^{i\hat{\varphi}/4}(c_\uparrow+c_\downarrow)\gamma_2\right. \nonumber \\ 
+ \left.e^{-i\hat{\varphi}/4}(c_\uparrow-c_\downarrow)\bar{\gamma}_3\right] + \mathrm{h.c},
 \end{align}
accounting for the hybridization of the Majorana fermions localized near the weak link, $\gamma_2$ and $\bar{\gamma}_3$, with the dot. We dub the combined model the Majorana-dot-Transmon (MDT). Together with the Majorana fermions $\bar{\gamma}_1$ and $\gamma_4$ at the nanowire’s remote ends \footnote{We assume negligible hybridization of the dot operators and the nearby $\gamma_2$ and $\bar{\gamma}_3$ with the Majorana fermions $\bar{\gamma}_1$ and $\gamma_4$ at the nanowire’s remote ends.}, they form the non-local fermions $\zeta_L=\frac{1}{\sqrt{2}}\left(\overline{\gamma}_1+i\gamma_2\right)$, $\zeta_R=\frac{1}{\sqrt2}\left(\overline{\gamma}_3+i\gamma_4\right)$. The coupling between the dot and the Majorana fermions is given by $w$.

\begin{figure}[b]
\includegraphics[width=1\linewidth]{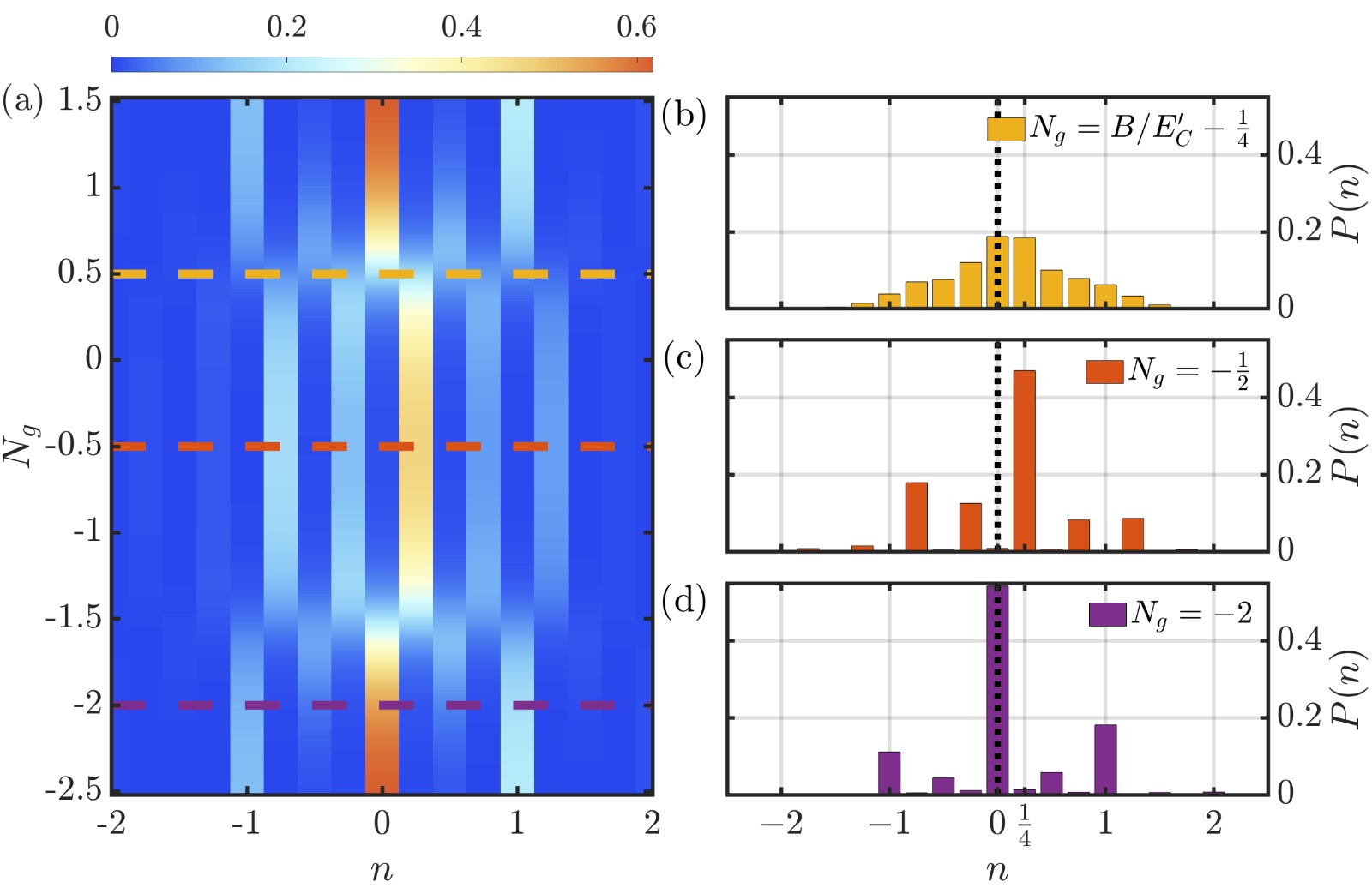}
 \caption{(a) Probability density $P(n)$ of the Majorana-Dot-Transmon's charge states $n$ as function of $N_g$. Three points of interest of $N_g$ were chosen and are marked by the horizontal dashed lines. The system transitions between two patterns of $P(n)$ at $N_g=-1.5$ and $N_g=0.5$ controlled by $V_g$.
 (b)-(d) show the probability density at specific values of $N_g$ that correspond in color to the dashed lines in (a). 
 The system parameters were set to $E_C=E_C'$, $E_C/E_J=7.14$, $\Gamma/E_C'=0.01$, $B/E_C'=0.75$, $w/E'_C=0.086$ and $n_g=1/10$.  }
\label{fig:Pn}
\vspace{-5mm}
\end{figure}

\begin{figure}[t]
\includegraphics[width=1.1\linewidth]{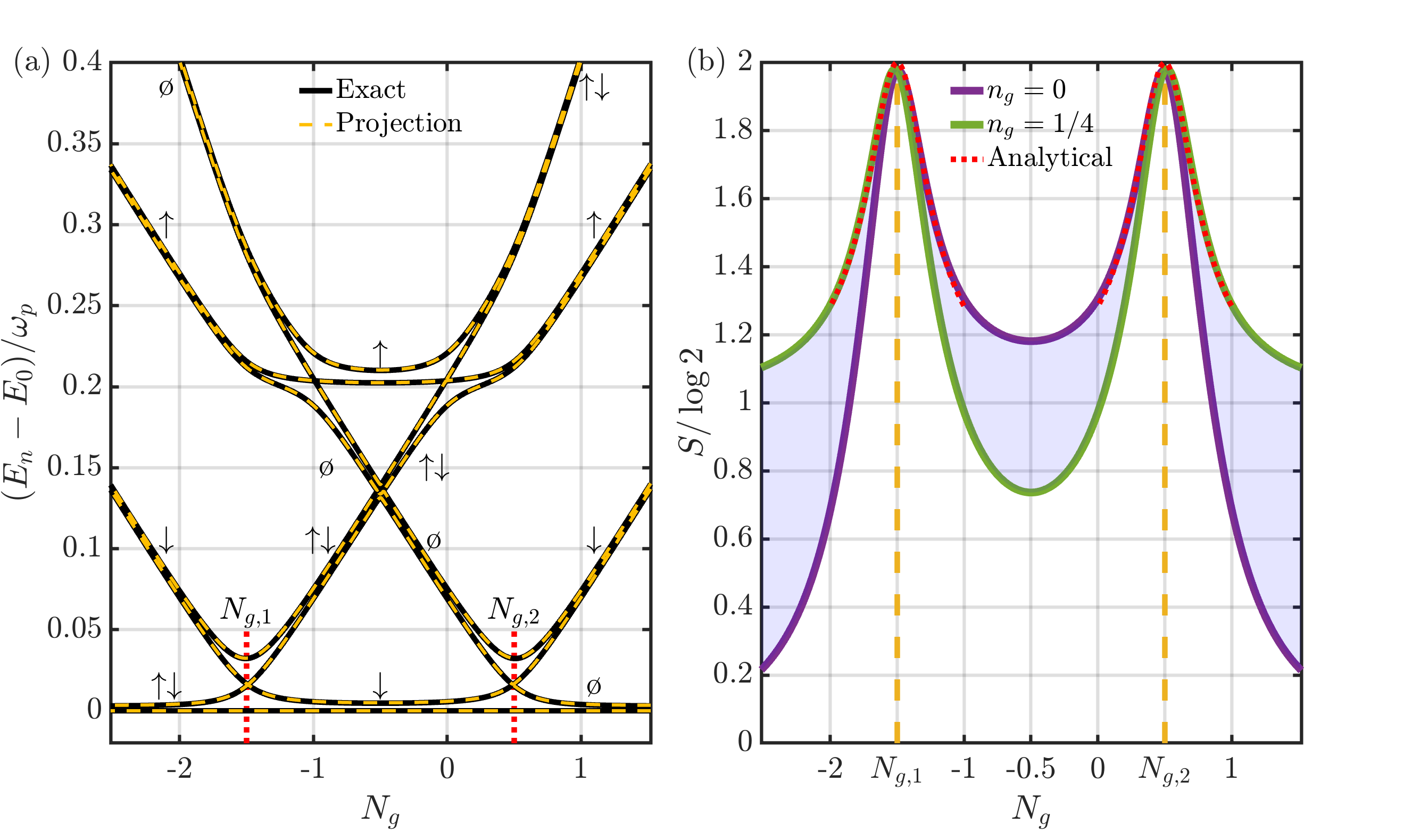}
 \caption{Spectrum characteristics of the Majorana-DT. (a) Fine-structure of the lowest Transmon level presented for the exact numerical analysis (solid line) and the projected 8-dimensional Hamiltonian (dashed line) with $n_g=1/10$. Throughout the figure, the approximate dot's spin sector is marked alongside its corresponding energy level. (b) The entanglement entropy $S$ as a function of the top gate $N_g$, with $n_g=0$ (purple) and $n_g=1/4$ (green). The blue shading represents $S$ for intermediate values of $n_g$. The red dotted line represents the approximate formula in Eq.~\eqref{eq:dmtr EE approx} which is independent of the side gate $n_g$. The peak points $N_{g,1}$ and $N_{g,2}$ are marked in dashed yellow lines. Here $N_t=0$, $E_C=E_C'$, $E_C/E_J=7.14$, $\Gamma/E_C'=0.01$,  $B/E_C'=0.75$ and $w/E'_C=0.086$.}
 \vspace{-5mm}
\label{fig: spectrum and entropy}
\end{figure}

To solve the model we extend the basis of the previous section to $|p_L,p_R,\sigma\rangle$, where $p_j=e,o$ indicate the parity on each side of the junction controlled by the occupation of $\zeta_j$, and $\sigma$ denotes the state of the dot. Since the total parity of the system is conserved, we shall consider the zero-parity case which consists of only eight possible fermion occupation subspaces. These are summarized in Table~\ref{eq: boundary conditions table - majorana dot Transmon}, where we use the notation $f_{p_L,p_R, \sigma}(\varphi)=\langle\varphi |p_L,p_R,\sigma\rangle$ for the wavefunctions in each subspace. The corresponding Hamiltonian,
\begin{align}
\label{eq: majorana-dot-Transmon  Hamiltonian matrix}
\mathcal{H}=\left(\begin{array}{cc}
    \mathcal{H}_1 & \mathcal{W} 	\\
    \mathcal{W}^\dagger &  	\mathcal{H}_2
\end{array}\right),
\end{align}
has eight subspaces which are ordered as in Table~\ref{eq: boundary conditions table - majorana dot Transmon}. The blocks on the diagonal $\mathcal{H}_1$ and $\mathcal{H}_2$ deal with the interaction between the Transmon and the dot, similar to the form in the previous section but with some added complexity due to the single particle occupation in the dot which also introduces the effect of the Zeeman field. The off-diagonal term $\mathcal{W}$ is responsible for the single fermion charge transfer via the Majorana-dot coupling (Eq.~\eqref{eq: majorana-dot-Transmon H_gamma}).

While a complete solution is daunting, here we are interested ultimately in groundstate physics so we will turn our attention to the lowest energy subspace, by projecting on the Transmon's ground state. This reduces the problem to an 8-dimensional Hamiltonian. The diagonal elements of $\mathcal{H}_1$ and $\mathcal{H}_2$ can be diagonalized by modifying the Transmon wavefunctions to comply with the boundary conditions of Table~\ref{eq: boundary conditions table - majorana dot Transmon}, with corresponding eigenenergies,
\begin{equation}
\label{eq: majorana dot Transmon - ground states}
\epsilon_{0,T} = \epsilon_0 + 
\left\{
\begin{aligned}
& (-1)^{(n_d + m_L + m_R)/2} t_0 \cos(2\pi n_g), & n_d \; \text{even}\\
& (-1)^{m_L} t_0 \sin(2\pi n_g), & n_d \; \text{odd}
\end{aligned}
\right.
\end{equation}
Here we introduced the notation $m_L, m_R$ for the occupation of the left and right non-local fermions respectively. The off-diagonal terms of Eq.~\eqref{eq: majorana-dot-Transmon  Hamiltonian matrix} can be found using the Harmonic approximation. This approach is valid if $\Gamma$ and $B$ are lower than the Transmon plasma frequency. 

We will now analyze the reduced density matrix of the charge states, including their diagonal elements (Fig.~\ref{fig:Pn}) and the entanglement entropy (Fig.~\ref{fig: spectrum and entropy}). Fig.~\ref{fig: spectrum and entropy}(a) demonstrates that the spectrum of the 8-dimensional projection discussed above agrees with exact diagonalization.

In Fig.~\ref{fig:Pn}, the probabilities $P(n)$ are displayed. The system undergoes transitions between different regimes at $N_g=-1.5$ and $N_g=0.5$ [Fig.~\ref{fig:Pn}(a)]. The intermediate pattern ($-1.5<N_g<0.5$) is associated with a single-electron occupation of the dot, while the other regimes are associated with an even number of electrons on the dot. In the transition points, four distinct sectors with offsets $0$, $1/4$, $1/2$ and $3/4$ are superposed [Fig.~\ref{fig:Pn}(b)].

The effect of the Majorana fermions is also apparent in the entanglement entropy. Following a projection on the Transmon's groundstate subspace the maximal value of the entropy, compared to the pure DT, increases to $S_\text{max} = \log(4)$. In Fig.~\ref{fig: spectrum and entropy}(b) we can see the entanglement entropy as a function of $N_g$, where the points corresponding to $S_\text{max}$ are given by $N_{g,1}=-B/E_C'-\frac{3}{4}$ and $N_{g,2} =B/E'_C-\frac{1}{4}$. These values indicate the points of avoided crossings and can be easily found by setting $\Gamma = w = t_0 = 0$ in the projected Hamiltonian. Near each of these points a slightly more general solution can be obtained by focusing on the $\Gamma, w, t_0 \ll B \ll \sqrt{8E_J E_C}$ regime, where the high magnetic field allows us to neglect the dot's $\uparrow$-states and the Hamiltonian is reduced to an easily diagonalizable one. Thus, near the points $N_{g,\ell}$ ($\ell = 1,2$), the ground state takes the form:

\begin{equation}
\begin{aligned}
|\text{gs}_\ell\rangle \simeq \frac{1}{\sqrt{2}} & 
\Big( \sin(\theta_\ell)|o,e,\downarrow\rangle - \sin(\theta_\ell)|e,o,\downarrow\rangle \\
- & \cos(\theta_\ell)|e,e,\sigma_\ell\rangle - \cos(\theta_\ell)|o,o,\sigma_\ell\rangle \Big),
\end{aligned}
\end{equation}
where $\sigma_1 = \uparrow\downarrow$, $\sigma_2 =\Emptyset$ and
\begin{equation}
\tan(2\theta_\ell) = (-1)^{\ell}\frac{2\sqrt{2} w \exp\left(-\frac{\varphi_0^2}{64}\right)}{E_C'(N_g-N_{g,\ell})}.
\end{equation}
We can see that indeed near the $N_{g,\ell}$ points the ground state approaches an equal superposition of four states and thus gives us the value $S_\text{max} = \log{(4)}$. To get the functional form of the entanglement entropy near the $N_{g,\ell}$'s points, we trace out all degrees of freedom (see Supplementary Material), except the implicit Transmon state, from the ground state density matrix $\rho_{\text{gs}_\ell} = |\text{gs}_\ell\rangle \langle \text{gs}_\ell|$. This gives us:
\begin{equation}\label{eq:dmtr EE approx}
\begin{aligned}
S_\ell =& -\cos^2(\theta_\ell) \log \cos^2(\theta_\ell) 
-\sin^2(\theta_\ell) \log \sin^2(\theta_\ell) \\ 
&+\log(2).
\end{aligned}
\end{equation}
Fig.~\ref{fig: spectrum and entropy}(b) shows a fairly good agreement between the exact numerical calculation of the entanglement entropy and the analytical approximation, valid only near the regions of the peaks. 

\emph{Conclusions.}--- Our investigation focuses on the ground state charge states of semiconductor-superconductor hybrids. We pinpoint distinctive features within the reduced density matrix of the charge states, particularly emphasizing low-lying interacting Andreev and Majorana states. The interacting Andreev states exhibit a characteristic distribution of relative charge numbers, $P(n)$, comprising both integers and half-integers. Conversely, the presence of Majorana fermions introduces a quarter-charge offset, resulting in a more intricate superposition of charge states, yielding four distinct sectors with offsets of $0$, $1/4$, $1/2$, and $3/4$. 

Moreover, we observe that the entanglement entropy associated with the charge states peaks when the dot's states become resonant with the condensate. This resonance manifests as a maximum entropy value of $\log(2)$ for a single Andreev state. Introducing Majorana fermions further enhances this effect, leading to peaks with a maximum entropy of $\log(4)$ for the Andreev state dressed by Majorana fermions. Importantly, in the latter scenario, the entanglement entropy peaks do not readily split into two distinct $\log(2)$ peaks, unlike the case of two simultaneously resonant Andreev states. 

\begin{acknowledgments}
E.Gi. and E.Gr. would like to thank the UUKi UK-Israel researcher mobility scheme for their support.
\end{acknowledgments}

\bibliography{tomography}

%%%%%%%%%%%%%%%%%%%%%%%%%%%%%%%%%%%%%%%%%%%%%%%%%%%%%%%%%%%%%%%%%%%%%%%%%%%%%%%%%%%%%%%%%%%%%%%%%%%%%%%
%\newpage

\setcounter{figure}{0}
\setcounter{equation}{0}
\makeatletter
\renewcommand{\thefigure}{S\@arabic\c@figure}
\renewcommand{\theequation}{S\@arabic\c@equation}
\makeatother

\appendix
\section{SUPPLEMENTARY MATERIAL}

\section{Tracing out degrees of freedom }

This section explains how we trace out degrees of freedom in our model and eventually calculate the entanglement entropy. In cases of a  numerical calculation, the wavefunction will be represented in the charge basis. The DT model has three quantum numbers: $N$ and $n$ which count the charge, and $\sigma$ for the dot states. The Majorana-DT introduces additional parity variations due the Majorana states, denoted by $p_L, p_R$ with their corresponding occupation $m_L, m_R$. As is in the body of the paper we constrain the total amount of cooper pairs $N$ by particle conservation $N_t=N+\frac{1}{2}n_d$ and take $N_t=0$.

\subsection{DT case}
\label{sec: appendix dot-Transmon}
As a single occupation of the dot will not mix with the Cooper pairs in this model, we are interested only in $\sigma\in \{ \Emptyset,\uparrow\downarrow\}$ subspaces. The charge difference $n$ has a selection rule that determines the possible values in each subspace: 
\begin{align}
n(\sigma)\in \mathbb{Z} + 
\begin{cases}
0, & \sigma = \Emptyset,\\
\frac{1}{2}, & \sigma = \uparrow \downarrow.
\end{cases}
\end{align}
 A pure state that obeys $N_t=0$ can be written as,
\begin{align}
|\psi\rangle = 
\sum_{\sigma\in \{\Emptyset,\uparrow\downarrow\}} \sum_{n(\sigma)}
\psi_{n(\sigma),N(\sigma),\sigma}|n(\sigma),N(\sigma),\sigma\rangle,
\end{align}
with $N(\sigma)=-n_d(\sigma)/2$.
Its corresponding density matrix is given by $\rho_{[n,N,\sigma]} = |\psi\rangle \langle\psi|$, and by tracing out the degrees of freedom $\sigma$ and $N$, $\rho_{[n]}=\sum _{N,\sigma}\langle N,\sigma| \rho_{[n,N,\sigma]}|N,\sigma \rangle$ it assumes a block-diagonal form, $\rho_{[n]}= 
\left(\begin{matrix}
\rho^{(0,\Emptyset)}_{[n\in \mathbb{Z}]} & 0 \\
0 & \rho^{(-1,\uparrow\downarrow)}_{[n\in \mathbb{Z}+\frac12]}
\end{matrix}\right)$, with each block $\rho_{[n]}^{(N,\sigma)}$ represented in the $n$-basis,
 \begin{align}
\rho^{(0,\Emptyset)}_{[n\in \mathbb{Z}]}&= 
\hspace{-0.2cm}\sum_{n,n'\in \mathbb{Z}}\hspace{-0.2cm}
\psi_{n,0,\Emptyset}\psi^*_{n',0,\Emptyset}|n\rangle
\langle n'|,
\\ 
\rho^{(-1,\uparrow\downarrow)}_{[n\in \mathbb{Z}+\frac12]}&=
\hspace{-0.4cm}\sum_{n,n'\in \mathbb{Z}+1/2}\hspace{-0.5cm}
\psi_{n,-1,\uparrow\downarrow}\psi^*_{n',-1,\uparrow\downarrow}|n\rangle
\langle n'|.
\end{align}
In the main article, we focus on two variations of Transmon's groundstate, which differ by their boundary conditions and thus their dependence on $n_g$, which we here denote $|\tau(\sigma)\rangle$:
\begin{equation}
\begin{aligned}
&\left(4E_C(\hat{n} - n_g)^2  -E_J \cos \hat{\varphi}\right) |\tau(\sigma)\rangle\\   &\qquad \qquad =\epsilon_{\text{T},0}(n_g+\tau(\sigma))|\tau(\sigma)\rangle,
\end{aligned}
\end{equation} 
where $\tau(\sigma)$ corresponds to the two subspaces:
\begin{align}
\tau(\sigma)=
\begin{cases} 
0, & \sigma = \Emptyset,\\
\tfrac{1}{2}, & \sigma=\uparrow \downarrow. 
\end{cases}
\end{align}
A pure state of the DT in the Transmon groundstate sector is given by
\begin{equation}
\begin{aligned}
|\psi\rangle &= 
\sum_{\sigma\in \{\Emptyset,\uparrow\downarrow\}}
\tilde{\psi}_{\tau(\sigma),N(\sigma),\sigma}|\tau(\sigma),N(\sigma),\sigma\rangle,
\end{aligned}
\end{equation}
Tracing out the degrees of freedom $\sigma$ and $N$ of its density matrix gives 
$\rho_{[\tau]}= 
\left( 
\begin{matrix}
\rho^{(0,\Emptyset)}_{[\tau]} & 0 \\
0 & \rho^{(-1,\uparrow\downarrow)}_{[\tau]}
\end{matrix}
\right)$,
where the blocks are
\begin{align}
\rho_{[\tau]}^{(0,\Emptyset)}&= 
|\tilde{\psi}_{0,0,\Emptyset}|^2|0\rangle\langle 0|,
\\ 
\rho_{[\tau]}^{(-1,\uparrow\downarrow)}&=
|\tilde{\psi}_{\frac{1}{2},-1,\uparrow\downarrow}|^2
|{\tfrac{1}{2}}\rangle 
\langle\tfrac{1}{2}|.
\end{align}
The entanglement entropy for this state is: 
\begin{equation}
\begin{aligned}
S=&-\mathrm{Tr}\rho_{[\tau]} \log \rho_{[\tau]}=-|\tilde{\psi}_{0,0,\Emptyset}|^2\log|\tilde{\psi}_{0,0,\Emptyset}|^2\\
& \qquad
-|\tilde{\psi}_{\frac{1}{2},-1,\uparrow\downarrow}|^2\log|\tilde{\psi}_{\frac{1}{2},-1,\uparrow\downarrow}|^2. 
\end{aligned}
\end{equation}
\\

\subsection{MDT case}
\label{sec: appendix majorana-dot-Transmon}

In this section, we extend the tracing-out procedure of the DT case to include the Majorana fermions. For simplicity we will use the notation $\kappa = (p_L, p_R)$ throughout the section. Here the charge difference $n$ has the selection rule: 
\begin{align}
n(\sigma,\kappa)\in \mathbb{Z}+
\begin{cases}
   \multirow{2}{*}{$\frac{1}{2} \left[\frac{n_d +m_L+m_R}{2}\text{ mod } 2\right]$,} 
                        & \sigma \in \left\{\Emptyset,\uparrow \downarrow\right\}\\
                        & \kappa\in \left\{(e,e), (o,o)\right\}\\
   \multirow{2}{*}{$\frac{1}{4}(m_L-m_R)$,} 
                        & \sigma \in \left\{\uparrow, \downarrow\right\}\\
                        &\kappa\in \left\{(o,e), (e,o)\right\}. \\
\end{cases}
\end{align}
A pure state that obeys $N_t=0$ can be written as
\begin{equation}
\begin{aligned}
|\psi \rangle  &= 
\hspace{-0.3cm}\sum_{\substack{\sigma\in \{e,e,\uparrow\downarrow\} \\  \kappa\in\{(e,e),(o,o)\}}}
\sum_{n(\sigma,\kappa)}
\psi_{n(\sigma,\kappa),N(\sigma),\sigma,\kappa}|n(\sigma,\kappa),N(\sigma),\sigma,\kappa\rangle
\\
&+\hspace{-0.3cm}\sum_{\substack{ \sigma\in \{\uparrow,\downarrow\}\\ \kappa\in\{(o,e),(o,e)\}}}
\sum_{n(\sigma,\kappa)}
\psi_{n(\sigma,\kappa),N(\sigma),\sigma,\kappa}|n(\sigma,\kappa),N(\sigma),\sigma,\kappa\rangle.
\end{aligned}
\end{equation}
Tracing out the dot degrees of freedom $\sigma$ and the total charge $N$ yields the reduced density matrix:
\begin{align}\label{eq:trace_out_dot}
\rho_{[n,\kappa]}=
\left(
\begin{array}{ccc}
\rho^{(0,\Emptyset)}_{[n,\kappa]}&0 &0 \\
0 &\rho_{[n,\kappa]}^{(-1,\uparrow \downarrow)}  &0 \\
0& 0 & \sum \limits_{\sigma\in \uparrow, \downarrow} \rho^{(-\frac12,\sigma)}_{[n,\kappa]}
\end{array}
\right).
\end{align}
Since $N+\frac{1}{2}n_d = 0$, the result is a block diagonal matrix. Each $\rho_{[n,\kappa]}^{(N,\sigma)}$ is spanned with $n$ and $\kappa$  states as follows:
\begin{align}
\rho_{[n,\kappa]}^{(0,\Emptyset)}&=
\hspace{-0.4cm}
\sum_{\substack{\kappa,\kappa'\in\{(e,e),(o,o)\}\\ n(\Emptyset,\kappa)}}\hspace{-0.45cm}
\psi_{n(\Emptyset,\kappa),0,\Emptyset,\kappa}\psi^*_{n'(\Emptyset,\kappa'),\Emptyset,\kappa'}
 \nonumber \\ 
& \qquad \times |n(\Emptyset,\kappa),\kappa\rangle\langle n'(\Emptyset,\kappa'),\kappa'|,\\
\rho_{[n,\kappa]}^{(-\frac12,\sigma\in\uparrow,\downarrow)}&=\hspace{-0.45cm}\sum_{\substack{\kappa,\kappa'\in\{(e,o),(e,o)\} \\ n(\sigma,\kappa)}}\hspace{-0.4cm}
\psi_{n(\sigma,\kappa),-\frac{1}{2},\sigma,\kappa}
\psi^*_{n'(\sigma,\kappa'),-\frac{1}{2},\sigma,\kappa'}
\nonumber\\
&\qquad \times |n(\sigma,\kappa),\kappa\rangle \langle n'(\sigma,\kappa'),\kappa'|,  \\
\rho_{[n,\kappa]}^{(-1,\uparrow\downarrow)}&= \hspace{-0.4cm}\sum_{\substack{\kappa,\kappa'\in\{(e,e),(o,o)\}\\ n(\uparrow\downarrow,\kappa)}}\hspace{-0.45cm}
\psi_{n(\uparrow\downarrow,\kappa),-1,\uparrow\downarrow,\kappa}
\psi^*_{n'(\uparrow\downarrow,\kappa'),-1,\uparrow\downarrow,\kappa'} \nonumber
 \\
 &\qquad \times |n(\uparrow\downarrow,\kappa),\kappa\rangle \langle n'(\uparrow\downarrow,\kappa'),\kappa'|.
\end{align} 
If we continue and trace over $\kappa$ the reduced density has a block diagonal form, and each block has different sets of $n\in \mathbb{Z}+[0,1/2,1/4,-1/4]$
\begin{widetext}
\begin{align}
\rho_{[n]}&=\sum_{\kappa}  \langle \kappa| \rho_{[n,\kappa]}|\kappa \rangle
= \left(
\begin{matrix}
\rho_{[n\in\mathbb{Z}]}^{(0,\Emptyset,e,e)}+\rho_{[n\in\mathbb{Z}]}^{(-1,\uparrow \downarrow,o,o)} & 0 & 0 & 0 \\
0 & \rho_{[n\in\mathbb{Z}+\frac12]}^{(0,\Emptyset,o,o)}+ \rho_{[n\in\mathbb{Z}+\frac12]}^{(-1,\uparrow \downarrow,e,e)} & 0 & 0\\
0 & 0 & \sum \limits_{\sigma\in \uparrow, \downarrow} \rho_{[n\in\mathbb{Z}+\frac14]}^{(-\frac12,\sigma,e,o)} & 0\\
 0&0 &0 &  \sum \limits_{\sigma\in \uparrow, \downarrow} \rho_{[n\in\mathbb{Z}-\frac14]}^{(-\frac12,\sigma,o,e)}
 \end{matrix}
\right).
\end{align}
\end{widetext}
As in the previous section, we focus on projected Transmon groundstate $|\tau(\sigma,\kappa)\rangle$, under the constraints given above the four projected $\tau(\sigma,\kappa)$ states, where: 
\begin{align}
\tau(\sigma,\kappa)=
\begin{cases}
   \multirow{2}{*}{$\frac{1}{2} \left[\frac{n_d +m_L+m_R}{2}\text{ mod } 2\right]$,} 
   & \sigma \in \{\Emptyset,\uparrow \downarrow\}\\
   & \kappa\in \{(e,e),(o,o)\}\\
   \multirow{2}{*}{$\frac{1}{4}(m_L-m_R)$,} 
   & \sigma \in \{\uparrow, \downarrow\}\\
   & \kappa\in\{(e,o),(o,e)\}\\
\end{cases}
\end{align}
a pure state in this subspace 
\begin{equation}
\begin{aligned}
|\psi \rangle  &= 
\hspace{-0.3cm}\sum_{\substack{\sigma\in \Emptyset,\uparrow\downarrow \\  
\kappa\in\{(e,e),(o,o)\}}}\hspace{-0.2cm}
\psi_{\tau(\sigma,\kappa),N(\sigma),\sigma,\kappa}|\tau(\sigma,\kappa),N(\sigma),\sigma,\kappa\rangle
\\
&+\hspace{-0.3cm}\sum_{\substack{ \sigma\in {\uparrow,\downarrow}\\ \kappa \in\{(e,o),(o,e)\}}}\hspace{-0.2cm}
\psi_{\tau(\sigma,\kappa),N(\sigma),\sigma,\kappa}|\tau(\sigma,\kappa),N(\sigma),\sigma,\kappa\rangle,
\end{aligned}
\end{equation}
likewise, we trace over $N$, $\sigma$ and $\kappa$ and get:
\begin{equation}\label{eq: EE dmtr}
\begin{aligned}
\rho_{[\tau]}=&\sum_{N,\sigma,\kappa}  \langle N,\sigma,\kappa| \rho_{[\tau,N,\sigma,\kappa]}| N,\sigma,\kappa \rangle \\
=& \left(|\psi_{0,0,\Emptyset,e,e}|^2+|\psi_{0,-1,\uparrow\downarrow,o,o}|^2 \right)
\left| 0\right\rangle\left\langle 0\right|\\
+&\left(|\psi_{\frac{1}{2},0,\Emptyset,o,o}|^2+|\psi_{\frac{1}{2},-1,\uparrow\downarrow,e,e}|^2\right)
\left|\tfrac{1}{2}\right\rangle
\left\langle \tfrac{1}{2}\right|\\
+&\left( |\psi_{\frac{1}{4},-1/2,\uparrow,o,e}|^2+|\psi_{\frac{1}{4},-1/2,\downarrow,o,e}|^2\right)
\left|\tfrac{1}{4}\right\rangle\left\langle \tfrac{1}{4}\right|\\
+&\left(|\psi_{-\frac{1}{4},-1/2,\uparrow,e,o}|^2+|\psi_{-\frac{1}{4},-1/2,\downarrow,e,o}|^2\right)
\left|-\tfrac{1}{4}\right\rangle
\left\langle -\tfrac{1}{4}\right|.
\end{aligned}
\end{equation}
and the entanglement entropy in this case $S=-\mathrm{Tr}\rho_{[\tau]}\log\rho_{[\tau]}$.

\end{document}